\documentclass{article}
\usepackage{amsfonts}
\usepackage{amsmath}

\begin{document}

\title{Comment on the paper "The finite square well: whatever is worth
teaching at all is worth teaching well ", \textit{arXiv}:1505.03376v2
[physics.ed-ph] by K. Razi Naqvi and S. Waldenstr\o m }
\author{ \and Victor Barsan \\
%EndAName
IFIN-HH, 30 Reactorului Str., Magurele-Bucharest, 077125, Romania}
\maketitle

\begin{abstract}
In the aforementioned paper, the authors claim that my results concerning
the consequent application of Garrett's method for obtaining approximate
expressions of the bound states energy of a particle in a finite rectangular
well are incorrect. I shall show hat this is not the case, and demonstrate
that both their and my results lead to the same conclusion, i.e. that the
consequent application of Garrett's method is equivalent to Barker's
approximation.
\end{abstract}

In \cite{Razi}, Razi Naqvi and Waldenstr\o m discuss the results published
in \cite{VB-RD} on the consistent application of Garrett's method for
evaluation of the energy of a particle in a finite square well \cite{Garrett}%
, claiming that they are incorrect. We shall explain that this is not the
case, and clarify in part the origin of Razi Naqvi and Waldenstr\o m's
confusion.

In fact, Garrett noticed that the energy of the $n-th$ bound state of a
particle in a finite square well can be approximated by the energy of the $%
n-th$ state of same particle, in a somewhat large infinite well; the width
of this infinite well is $n-$dependent. Garrett used this remark to get "a
simple iterative approximation for the energy states of the finite well"; he
applied this scheme in a two-step iteration. Our contribution in \cite{VB-RD}
was to apply Garrett's approach in an iteration with an infinite number of
steps, and our main result is that, in this way, the energy of the $n-th$
bound state of a particle in a deep finite well, for small values of $n,$
written as a power series in the inverse of the potential strength,
coincides with the Barker approximation.

Some of the results from \cite{VB-RD} have been recently presented and
expanded in a form more suited to teaching quantum mechanics at an
elementary level in \cite{VB-EJP}. This is why I shall discuss the
objections of Razi Naqvi and Waldenstr\o m to our results using their most
recent form, i.e. \cite{VB-EJP}.

Razi Naqvi and Waldenstr\o m make a detailed analysis of the numerical
example discussed by Garrett \cite{Garrett}, notice that their accuracy is
quite modest, and conclude that "Garrett's \textit{scheme} does not provide
useful results". We are also doing an analysis of the same numerical
example, which does not actually correspond to a deep well (the potential
strength is $P=4,$ so the condition for a deep well, $P\gg 1$ is hardly
fulfilled), but we focus on the physical significance of the errors (\cite%
{VB-EJP}, Sections 5 and 6). As far as we obtain the equivalence between
Garrett's approach (consistently applied, i.e. with an infinite number of
iterations) and Barker approximation, we do not discuss anymore its
accuracy, as Barker approximation is well understood (in fact, the formula
giving this approximation is extremely simple (see eq. (34) in \cite{VB-EJP}%
) and already used in several papers of applied physics (see for instance
ref. [16] in \cite{VB-EJP}).

However, Razi Naqvi and Waldenstr\o m consider that Garrett's idea "can be
placed on a sound basis", replacing the $n-$dependent infinite well
associated to the $n-$th level of the finite well with a unique infinite
well, having a width conveniently chosen. More exactly, for a finite well of
width $L,$ one can choose a quantity $\Delta $ so that the energy of the $n-$%
th state of the infinite well of width $L+\Delta $ coincides with the energy
of the $n-$th state of the finite well, in the Barker approximation.

So, Razi Naqvi and Waldenstr\o m re-obtain our result, claiming that it is
incorrect! Let us explain, in part, the origin of their conclusion. The
expansion of the ratio $k_{n}/k_{n}^{\left( \infty \right) }$, where $k_{n}$
is the wave vector corresponding to the $n-th$ bound state in the finite
well (it corresponds to the roots of the equation (37) of \cite{VB-EJP} for $%
k$), and $k_{n}^{\left( \infty \right) }$ - to the $n-th$ state of an
infinite well having the same width, in powers of $p=1/P,$ was given in \cite%
{PhilMag2014}, eq. (52). The first two terms of this expansion,
corresponding to the Barker approximation, do not depend on $n,$ according
to eqs. (53-55) of \cite{PhilMag2014}; this is why the $n-$dependence of the
energy of the $n-th$ bound state in a finite well has the same form.
However, the cubic and all the other higher order terms are $n-$dependent.
In fact, Razi Naqvi and Waldenstr\o m's main conclusion is that, with a
convenient choice (of the parameter $\Delta $ in eq. (35) of \cite{Razi},
the one-iteration Garrett's scheme is equivalent to Barker approximation.
This conclusion is correct, but the authors give no explanation for the fact
that the same equivalence cannot be extended to higher orders. Let us
mention also that, contrary to the Razi Naqvi and Waldenstr\o m's opinion,
our quantity of $\delta _{n}^{\left( \infty \right) }$ is both $n-$ and $p-$%
dependent, as a root of eq. (21) of \cite{VB-EJP}, but the $n-$dependence
drops in the approximation used in solving this equation.

A final remark: Razi Naqvi and Waldenstr\o m give an incorrect relation for
the number of roots of the eigenvalue equations for the wavevector (or
energy) of a bound state, taking into consideration, for their conclusion,
the extremum points of the functions $\sin x,\ \cos x,$ instead of $\sin
x/x,\ \cos x/x.$ The correct relation is given in \cite{PhilMag2014}, eq.
(28) and the subsequent discussion.

\bigskip

\end{document}